\documentclass[apjl]{emulateapj}

\usepackage{graphicx}
\usepackage{dcolumn}
\usepackage{bm}

\newcommand{\Dmax}{D_\mathrm{max}}

\newcommand{\lss}{\lambda}

\newcommand{\rg}{$r_g$}
\newcommand{\mfp}{$\lambda$}
\newcommand{\Lxp}{\lambda(x,p)}

\newcommand{\Kol}{Kolmogorov}

\newcommand{\res}{resonant}
\newcommand{\nonres}{nonresonant}
\newcommand{\xx}[1]{\!\times\!10^{#1}}

\newcommand{\xfeb}{x_\mathrm{feb}}

\newcommand{\rgzero}{r_{g0}}
\newcommand{\rgone}{r_\mathrm{g1}}
\newcommand{\Beff}{B_\mathrm{eff}}

\newcommand{\Lturb}{L}
\newcommand{\kdiss}{k_d}
\newcommand{\kmps}{km s$^{-1}$}
\newcommand{\NL}{nonlinear}

\newcommand{\Lfeb}{x_\mathrm{feb}}
\newcommand{\alf}{Alfv\'en}

\newcommand{\SCly}{self-consistently}
\newcommand{\Rtot}{R_\mathrm{tot}}

\newcommand{\muG}{$\mu$G}

\newcommand{\pmax}{p_\mathrm{max}}
\newcommand{\Emax}{E_\mathrm{max}}
\newcommand{\be}{\begin{eqnarray}}
\newcommand{\ee}{\end{eqnarray}}

\newcommand{\syn}{synchrotron}

\newcommand{\growthrate}{\gamma_\mathrm{nr}}
\newcommand{\lcor}{l_\mathrm{cor}}

\newcommand{\kstar}{k_\mathrm{*}}

\shorttitle{Nonresonant Instability in SNR shocks}
\shortauthors{Vladimirov, Bykov, \& Ellison}

\begin{document}

\title{Spectra of magnetic fluctuations and relativistic particles produced by a nonresonant wave instability
in supernova remnant shocks}

\author{Andrey E. Vladimirov}
  \affil{North Carolina State University, Department of Physics, Raleigh, NC 27695-8202, USA}
  \email{avladim@ncsu.edu}
\author{Andrei M. Bykov}
  \affil{Ioffe Physical-Technical Institute, St. Petersburg, Russia}
  \email{byk@astro.ioffe.ru}
\author{Donald C. Ellison}
  \email{don\_ellison@ncsu.edu}
  \affil{North Carolina State University, Department of Physics, Raleigh, NC 27695-8202, USA}

\date{\today}

\begin{abstract}
We model strong forward shocks in young supernova remnants with
efficient particle acceleration where a nonresonant instability
driven by the cosmic ray current amplifies magnetic turbulence in
the shock precursor. Particle injection, magnetic field
amplification (MFA) and the nonlinear feedback of particles and
fields on the bulk flow are derived consistently. The shock
structure depends critically on the efficiency of turbulence
cascading.  If cascading is suppressed, MFA is strong, the shock
precursor is stratified, and the turbulence spectrum contains
several discrete peaks.  These peaks, as well as the amount of MFA,
should influence \syn\ X-rays, allowing observational tests of
cascading and other assumptions intrinsic to the nonlinear model of
\nonres\ wave growth.

\end{abstract}


\section{Introduction}

Diffusive shock acceleration (DSA) in supernova remnant (SNR) shocks is
the most likely mechanism for the origin of the bulk of galactic cosmic
rays (CRs) up to at least $10^{15}$\,eV.
To produce the observed density of galactic CRs, the acceleration
mechanism must be efficient and, therefore, nonlinear
\citep[e.g.,][]{ BE1987, MD2001, Hillas2005}.
X-ray imaging of young SNRs reveals thin \syn\ filaments that are
best explained as originating from shock accelerated TeV electrons
radiating in strongly amplified circumstellar magnetic fields
\citep[see ][and references therein]{VL2003,
BambaEtal2003,Uchiyama_J1713_2007, Vink2008}.
These fields are almost certainly produced as part of
the DSA process \citep[e.g.,][]{BL2001, Bell2004, AB2006,
PLM2006, VEB2006, ZP2008, MD2009, Riquelme2009}.

\citet{Bell2004} suggested that magnetic fluctuations in a shock
precursor may be amplified by a fast, nonresonant instability induced by
the strong diffusive electric current of the shock-accelerated CRs.  The
transverse MHD waves produced by this mechanism have wavelengths much
shorter than the wave generating CR gyroradii.  In efficient DSA,
the CR current driving the instability must be strongly coupled to the
turbulent magnetic fields and supersonic MHD flow.
To address this issue, we simulated the amplification of Bell's modes as
part of a nonlinear Monte Carlo (MC) model of DSA \citep{VEB2006}.
We \SCly\ model four basic strongly coupled components of the
system: the bulk plasma flow, the full particle spectrum, the
self-generated MHD turbulence including cascading, and thermal
particle injection.
The efficiently produced CRs modify the upstream flow speed, and the CR
current determines the growth of stochastic magnetic fields. The
fields, in turn, set the particle diffusive transport properties and,
subsequently, the injection and acceleration efficiency of the
particles, closing the system.

\section{The Model}

Our Monte Carlo model of \NL\ particle acceleration with strong,
resonant, MFA is presented in \citep[][]{VEB2006, VBE2008}.
Here, we replace \res\ growth with the fast \nonres\ instability of
Bell and
  include the cascading of that turbulence to shorter wavelengths. The
  momentum and space dependent particle mean free paths are calculated
  consistently with the turbulence.

Consider the nonlinear precursor of a strong, plane-parallel,
steady-state shock. In the reference frame co-moving with the subshock
(at $x=0$), where the plasma flows in the positive $x$-direction, the
flow speed $u(x)$ has the value $u_0$ far upstream. The speed, $u(x)$,
drops in the precursor until it obtains the downstream speed
$u_2=u_0/\Rtot$, where $\Rtot$ is the overall shock compression ratio.
The nonresonant current instability
is assumed to be the source of
the wave spectrum energy density $W(x,k)$ (here $k$ is the wavenumber,
and $W dk$ is the amount of energy in the interval $dk$ per unit
spatial volume)
according to the following equation:
\begin{equation}
\label{eq_turb_evol}
u\frac{\partial W}{\partial x} + \frac{\partial \Pi}{\partial
  k} =  \growthrate W - \Lturb.
\end{equation}
The quantity $\Pi(x,k)$ describes cascading, i.e., the transfer of
turbulence energy from long to short wavelengths, $\Lturb(x,k)$
represents the dissipation of turbulence, and $\growthrate(x,k)$ is the
quasi-linear rate of wave energy amplification by the instability, and
is given by (see \cite{Bell2004}),
\begin{equation}
\label{bell_increment} \growthrate =  2 v_A
k\sqrt{\displaystyle\frac{k_c}{k}-1}, \quad
  \mathrm{for} \quad 1/\rgone < k < k_c .
\end{equation}
Here $v_A(x)=B_0/\sqrt{4 \pi \rho(x)}$ is the \alf\ speed, $c$ is the
speed of light, $B_0$ is the far upstream magnetic field directed
towards the shock normal, $\rho(x)$ is the thermal plasma mass density,
$\rgone(x)$ is the gyroradius of the least energetic current generating
CR, the critical wavenumber $k_c(x) = 4\pi j_d(x)/(c B_0)$, and the local
diffusive electric current of CRs responsible for the instability,
$j_d(x)$, is determined by the MC simulation.

Bell's derivation of (\ref{bell_increment}) assumes that the gyroradii
of the streaming CRs are much greater than the wavelengths of
the generated waves, which is expressed here by the condition $1/\rgone
< k$.
The growth rate, $\growthrate$, has a maximum at $k=k_c/2$ and
vanishes for $k>k_c$.

Nonlinear interactions between turbulent harmonics may lead to
  a re-distribution of the energy of turbulent fluctuations in $k$-space
  (i.e., spectral energy transfer). 
  The term $\partial \Pi/ \partial k$ in
  Equation~(\ref{eq_turb_evol}), with $\Pi>0$,
  describes a transfer of energy from large to 
  small turbulent fluctuations (i.e., cascading).
  Such a description, i.e., the Kolmogorov cascade model,
  has been successful in explaining the spectra of locally
  isotropic, incompressible turbulence in non-conducting fluids,
  observed in experiments and simulations
  \citep[e.g.,][]{Biskamp2003}.
  However, it is uncertain how spectral transfer operates in a collisionless shock
  precursor with a CR current strong enough to modify the MHD 
  modes, and in the presence of strong magnetic turbulence.
  In weak MHD turbulence, cascading was
  shown to be anisotropic \citep[][]{GS95}: harmonics with wavenumbers
  transverse to the mean magnetic field experience a Kolmogorov-like
  cascade, while the cascade in wavenumbers parallel to the mean field is
  suppressed. 
  \cite{DM2007} argue for an inverse cascade (i.e., from short to long
  wavelengths) of turbulent harmonics in the context of magnetic field
  amplification in DSA.
  Spectral energy transfer
  in the form of anisotropic diffusion of energy in $k$-space
  was proposed by \cite{Matthaeus2009}.
  MHD modeling of Bell's instability with fixed CR current, performed
  by \citet{Bell2005} and \citet{ZPV2008}, show that the magnetic
  structures consist of expanding walls of strong magnetic field
  spiraling around cavities in density and magnetic field. Moreover,
  \citet{ZPV2008} revealed that the turbulent energy is non-linearly
  transferred to both longer and shorter scales (see their Figure 4).
  It should be noted however, that the MHD models ignored both the
  interaction of CR particles with the instability-generated turbulence,
  and the feedback of the turbulence on the CR current that are the main
  subjects of our study.  
  The deflection of CRs by magnetic fluctuations
  may lead to important nonlinear effects, such as a saturation of the
  instability \citep[e.g.,][]{Bell2005, PLM2006}. 
  A complete numerical
  description of turbulence, that includes the evolution of the fields and
  the motion of the particles, with a realistic dynamic range, is
  currently unfeasible. 
  Because the primary goal of our present research
  is to study the non-linear aspects of DSA with Bell's instability
  including the interaction of CR particles with the instability-generated
  turbulence, approximations must be made for the poorly known details of
  the turbulent spectral energy transfer. 
  In this work, we investigate the consequences of different spectral
  energy transfer regimes by presenting two limiting cases. 
  In one (shown with solid curves in all plots), cascading is fully
  suppressed, i.e., $\Pi=0$. In the other (shown with dotted curves in all
  plots), the cascading from large to small scales is efficient and
  has the differential form \citep[see, e.g.,][]{Verma2004}, $\Pi =
  W^{3/2}k^{5/2} \rho^{-1/2}$, corresponding to Kolmogorov's 
  model\footnote{Note that we ignore a constant on the order of unity in
  this expression \citep[e.g.,][]{Verma2004}.}.

The dissipation term, $\Lturb$, is assumed to be zero with no cascading
and to have the form of viscous dissipation, $\Lturb =
v_Ak^2\kdiss^{-1}W$, in the model with cascading \citep[e.g.,][]{VBT93}.
The wavenumber, $\kdiss$, is identified with the inverse of a thermal
proton gyroradius: $\kdiss(x) = eB_0/[c\sqrt{m_pk_B T(x)}]$, where $m_p$
is the proton mass, $k_B$ is Boltzmann's constant and $T(x)$ is the
local gas temperature determined from the heating induced by $\Lturb$,
as described in \cite{VBE2008}. We will consider elsewhere
  models of spectral transfer different from Kolmogorov cascading.

To highlight the effects of the Bell nonresonant instability and
  cascading in strong SNR shocks of SNRs, we here
  ignore resonant instabilities and the compression
  of turbulence \citep[see][]{VEB2006}, even though these effects may
  play an important role in weaker shocks \citep[see][]{PLM2006}. We
  have performed more complete calculations including resonant and
  nonresonant instabilities with compression and cascading (to be
  demonstrated elsewhere) and confirmed that the qualitatively new
  findings presented here stand out even when resonant instabilities and
  compression are included.

Equation~(\ref{eq_turb_evol}) is integrated with respect to $x$
using the far upstream ($x=-\infty$) boundary condition $W\propto
k^{-1}$.
We assume that these far upstream seed fluctuations are linear waves
that are not subject to cascading or dissipation, and that the
transition to the turbulent regime takes place at a
position $x_t$ where the amplified wave spectrum
reaches the value $kW(x_t,k) = B_0^2/4\pi$ at some $k$. At this point,
in the model with cascading, $\Pi$ and $\Lturb$ are set from zero to the
values defined above.

The mean free path of a particle with momentum $p$, $\Lxp$, is
determined by the MHD turbulence. We calculate \mfp\
by combining well known theories in
different parameter ranges. For the highest energy particles with
gyroradii, \rg, that exceed the turbulence correlation length $\lcor$,
we assume $\lambda=r_g^2/\lcor \propto p^2 $, as described in
\cite{Toptygin1985} (see also \cite{Jokipii1971}).  Particles with
\rg\ that resonate with the field fluctuations are
assumed to have $\lambda = r_g / \mathcal{F}$ \citep[e.g.,][]{BE1987},
where $\mathcal{F}$ is the normalized product $k W$ evaluated at $k$
resonant with the particle momentum $p$.
The lowest energy particles, with $r_g \ll \lcor$, are assumed to be
trapped by turbulent vortices, and their diffusion is determined by the
dynamics of these vortices; these particles are assumed to have
$\lambda=\lcor$, independent of $p$ (e.g.,
\cite{Palmer1982}).
Our model for $\lambda$ smoothly interpolates between these regimes by
introducing a wavenumber $\kstar$ that, for a given particle
momentum $p$, separates the turbulence spectrum $W$ into the
small-scale ($k>\kstar$) and the large-scale ($k<\kstar$) parts. In
addition, we assume that thermal particles not yet injected into the
acceleration process have the Bohm mean free path $\lambda = r_g$. Full
details of the model are given in \cite{Vladimirov_Diss}.

The diffusive CR current $j_d(x)$ is calculated as the corresponding
moment of the accelerated proton distribution function $f(x,{\bf
p})$ derived by the MC simulation.

At every upstream location $x$, the minimal CR momentum $p_1$ is
defined from the particle distribution function,
$f(x,{\bf p})$, as the momentum below which the CRs contribute $1\%$ of
the total CR pressure. Then $p_1$ is used to calculate the gyroradius
$\rgone$ that determines the applicability range
of~(\ref{bell_increment}). We assume $\growthrate=0$ outside of that
range of $k$.

\section{Results}

For both examples shown here, we model a shock of speed
$u_0=10^4$~\kmps\ propagating in a uniform magnetic field $B_0=3$~\muG\
in a plasma with proton density $n_0=0.3$~cm$^{-3}$ and temperature $T_0
= 10^4$~K.  We assumed that the acceleration process is in steady-state
and size-limited with a free escape boundary at $\Lfeb =-10^7\,\rgzero$,
where $\rgzero \equiv mu_0c/eB_0 \approx 3.5\xx{10}$\,cm.
Note that $\pmax$ scales with $\xfeb$. In an actual shock, the
finite size, and/or finite age, of the shock will determine $\pmax$. 

\begin{figure}[ht]
  \includegraphics[angle=0, width=3.375in]{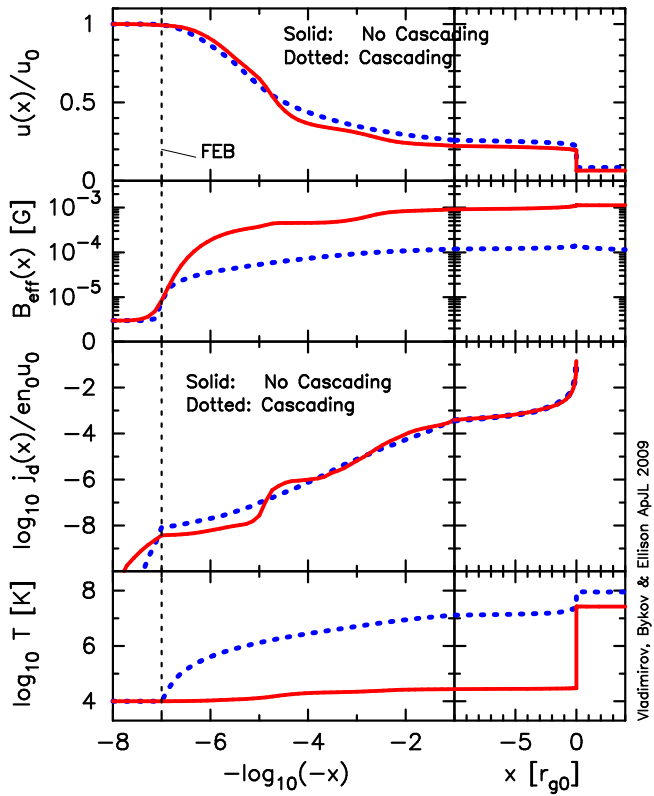}
\caption{Flow speed $u$, effective magnetic field $\Beff$, diffusive CR
  current $j_d$, and temperature $T$ vs. position $x$. Solid curves are
  without cascading and dotted curves include strong \Kol\
  cascading. The viscous subshock is at $x=0$ and the position of the
  free escape boundary (FEB) is indicated.
  \label{fig_ubj}}
\end{figure}

In Fig.~\ref{fig_ubj} we compare the flow speed, $u(x)$, the
amplified effective field, $\Beff(x)$, the CR current, $j_d(x)$, and
the thermal plasma temperature, $T(x)$, for the two models.
With no cascading (solid curves),
the overall compression ratio
$\Rtot=u_0/u_2 \approx 15$,
and the downstream field, $\Beff(x>0) \approx 1000$\,\muG.
These quantities are substantially reduced with cascading (dotted
curves), i.e.,
$\Rtot \approx 11$,
and $\Beff(x>0) \approx 100$\,\muG.
There are also differences in $j_d(x)$, where $j_d$ is smooth with
cascading but shows an uneven structure without cascading. The parameter
that determines the scale of the generated turbulence, $k_c$, has the
same $x$-dependence as $j_d$.  This uneven precursor structure has a
striking effect on the magnetic turbulence (see Fig.~\ref{fig_wp}).
Another prominent difference is the significantly increased precursor
temperature, $T(x)$, with cascading due to the dissipation of turbulence
at large $k$ (see \cite{VBE2008}).

\begin{figure}[ht]
\includegraphics[angle=0, width=3.375in]{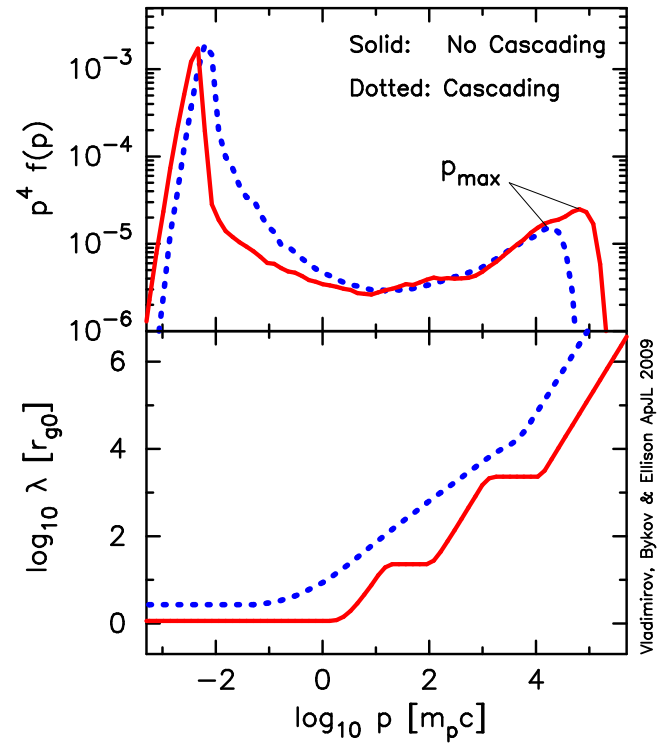}
\caption{Proton distribution function, $f$, times $p^4$, and
         accelerated particle mean free path, $\lambda$, both calculated
         downstream of the subshock. The turnover momenta, $\pmax$, for
         the two cascading models are indicated.
  \label{fig_fl}}
\end{figure}

In Fig.~\ref{fig_fl} we plot the downstream ($x=5 \, \rgzero$)
particle distribution $p^4 f(x,p)$ and mean free path
$\lss(x,p)/\rgzero$.
The maximum momentun, $\pmax$, is clearly greater without cascading
($\sim 10^5$ vs. $\sim 2\xx{4}\, m_p c$) but the concave shape for $p^4
f$ is present in both cases, indicating that both are efficient particle
accelerators \citep[e.g.,][]{BE99}.
As indicated by the thermal peaks, the shocked temperature is
considerably higher in the model with cascading, an effect similar to
what was observed in \cite{VBE2008}.
With cascading (dotted curves), $\lss$ is a smooth function of
$p$ showing the three regions discussed above: $\lss \sim$ constant for
low $p$, $\lss \propto p$ for intermediate values of $p$, and $\lss
\propto p^2$ for the highest momenta. Without cascading, $\lss$ is
always less than with cascading and shows plateau regions where
particles are trapped by vortices of different scales.

\begin{figure}[ht]
\includegraphics[angle=0, width=3.375in]{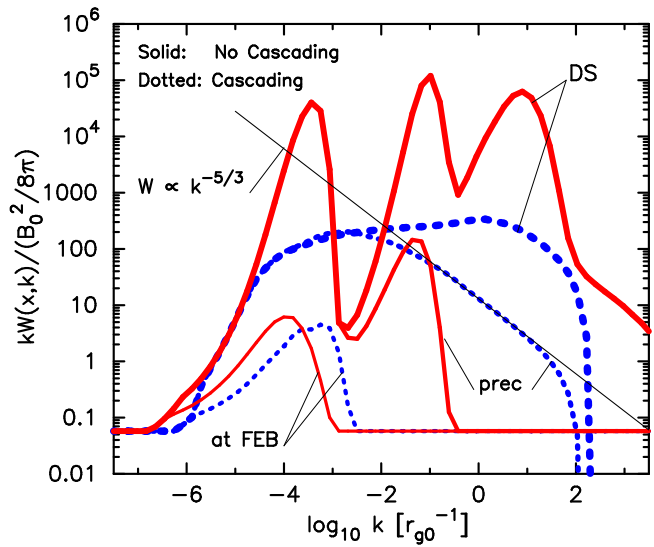}
\caption{Turbulence spectra, $W$, times $k$, at different locations
relative to the subshock. The lightest-weight curves are calculated at
$\Lfeb = -10^7\,\rgzero$, the heaviest curves are calculated in the
downstream region, and the medium-weight curves are calculated in the
precursor at $x = -1.2\xx{4}\,\rgzero$. The far upstream seed turbulence
(at all $k$) is at the level indicated by the horizontal lines.
  \label{fig_wp}}
\end{figure}

\section{Discussion}

Our most intriguing result is shown in Fig.~\ref{fig_wp} where we
plot $k W(x,k)$ at different locations, $x$, relative to the
subshock, as indicated. The seed turbulence from the unshocked
interstellar medium is at the level of the horizontal lines.

Without cascading (solid curves), we see fairly narrow peaks centered at
distinct $k$. This is in stark contrast to the case with cascading
(dotted curves), and to the broad spectrum of $kW(x,k)$ seen previously
with \res\ interactions \citep[e.g.,][]{VEB2006}.
Only the most energetic CRs reach a far upstream position near $\Lfeb$,
and the small current these CRs produce gives
a maximum $\growthrate$ at small $k = k_c/2$.
At $\Lfeb=-10^7\,\rgzero$, where the light-weight curves are calculated
in Fig.~\ref{fig_wp}, $\growthrate$ is a maximum at $k\sim
1\xx{-4}\,\rgzero^{-1}$ with no cascading, and at $k\sim
5\xx{-4}\,\rgzero^{-1}$ with cascading.
The difference corresponds to the difference in escaping CR flux at
$\Lfeb$.
At a position within the precursor at $x=-1.2\xx{4}\,\rgzero$
(medium-weight curves), a second peak at larger $k$ has developed when
cascading is suppressed. At this position closer to the subshock, lower
energy particles appear in large numbers and the generation of waves at
$k\sim 10^{-4}\,\rgzero$ shuts down because of the range in $k$ given in
Eq.~\ref{bell_increment}. The increased number of lower energy
particles, and the accordingly increased $j_d$, now corresponds to a
greater $k_c$ and shorter wavelength structures get amplified at
$k \sim 10^{-1}\;\rgzero^{-1}$.
The position of the leftmost peak is nearly constant between $\Lfeb$ and
$-1\xx{5}\,\rgzero$ because $j_d$ is nearly constant in this range, as
indicated in Fig.~\ref{fig_ubj}. At the downstream position (heaviest
solid curve), three distinct peaks are generated without cascading.

The same turbulence generation model with cascading shifts energy from
small to large $k$ and produces smooth spectra. Far enough upstream in
the precursor but not at $\Lfeb$ (medium-weight dotted curve), where
only relatively long wavelengths are amplified, the spectrum exhibits
the ``injection range'' at $k \lesssim 10^{-3} \; \rgzero^{-1}$, the
broad ``inertial range'' $10^{-3} \, \rgzero^{-1} < k < 10\,
\rgzero^{-1}$ with the Kolmogorov spectrum $W \propto k^{-5/3}$, and the
``dissipation range'' at $k \gtrsim 10\,\rgzero$.
Closer to the subshock, the increased $j_d$ shifts turbulence
amplification to greater $k$ making the spectrum harder. The downstream
spectrum becomes close to $W \propto k^{-1}$ over a broad range in
$k$.

The peaks occur because of the coupling of particle transport with
turbulence amplification. The first (smallest $k$) peak
forms far upstream, where only the highest energy particles are present,
and their current $j_d$ is low.  These particles diffuse in the
$\lambda\propto p^2$ regime, scattered by the short-scale magnetic field
fluctuations that they themselves generate. As the plasma moves toward
the subshock, advecting the turbulence with it,
lower energy particles appear. At some $x$, particles with energies low
enough to resonate with the turbulence generated farther upstream (in
the lowest $k$ peak) become dominant.  This strong \res\ scattering
leads to a high gradient of $j_d$ (seen at $x \sim -10^5\;\rgzero$ in
the third panel of Fig.~\ref{fig_ubj}), and the wavenumber $k_c/2$, at
which the amplification rate $\growthrate$ has a maximum, increases
rapidly.
The increased value of $k_c/2$ leads to the emergence of the second peak
between $10^{-2}$ and $10^{-1}\,\rgzero^{-1}$, as seen in
Fig.~\ref{fig_wp}.  Similarly, the third peak is generated at distances
closer to the subshock than $\sim -1\xx{4}\,\rgzero$ and this is seen in
the heavy-weight, downstream spectrum in Fig.~\ref{fig_wp} at $k \sim
10\,\rgzero^{-1}$ \footnote{The tail at $k > 100\,\rgzero^{-1}$ is
produced by the large $j_d$ at $-1\,\rgzero < x < 0$ seen in
Fig.~\ref{fig_ubj}.  The short-scale turbulence produced by this current
contains little power and does not impact our results for larger scales.}.

The number of peaks depends on the dynamic range, i.e., on $\Lfeb$. A
smaller $\Lfeb$ can result in two peaks, while a larger $\Lfeb$, and
therefore a larger $\pmax$, can yield four or more peaks in the
downstream region. As mentioned above, $\xfeb$ is a parameter in
our model and is chosen here to give $\Emax\sim
10-100$\,TeV.

The formation of the spectrum with discrete peaks occurs simultaneously
with the stratification of the shock precursor into layers (see the
plots of $j_d$), in which vortices of different scales are formed. The
peaks are a direct result of Bell's \nonres\ instability, but they will
not show up unless $\lss$ and $\growthrate$ are calculated consistently,
and the simulation has a large enough dynamic range in both $k$ and $p$.

As emphasized above, cascading in strong
turbulence is uncertain and our two models bound the extremes of no
cascading and \Kol\ cascading to larger $k$. Our results are
  also limited by the fact that we haven't
  considered the case of energy transfer from short to long
  scales. Nevertheless, the
stark differences we see in both the
shock structure and the turbulence spectrum suggest that observations
may be able to constrain cascading models.  The stratification is
eliminated and the peaks are spread out if rapid Kolmogorov cascading is
assumed. Downstream, the spectrum becomes close to, but slightly flatter
than $W\propto k^{-1}$.  Furthermore, the amplified effective magnetic
field, the shock compression ratio, and the maximum energy of
accelerated particles are all significantly smaller with cascading.

The large amount of energy observed in the first (smallest $k$) peak
of the turbulence spectrum without cascading defines a potentially
observable spatial scale. For the parameters used here, $\Dmax
\approx 2\pi/k \approx 2\xx{4}\,\rgzero \approx
10^{15}\;\mathrm{cm}$, but this depends on $\Lfeb$, and we have
found (work in preparation) that $\Dmax \propto |\Lfeb| \propto
\pmax$. Therefore, spatially resolved intensity variations translate
to $\Dmax$ \citep[see][]{BUE2008} and may offer a new way to
estimate $\pmax$ from X-ray \syn\ observations of SNRs.
For our parameters, the turbulence on the scale $\Dmax$
has the effective magnetic field $\Delta B \approx \sqrt{ 4 \pi k W }
\approx 400$~\muG, and the corresponding \alf\ speed $v_A \approx
400$~\kmps, giving a characteristic
time scale $t=\Dmax/v_A\approx 0.5$~yr. A larger
$\pmax$ implies a larger $t$.
This rapidly varying field may influence both the spectral shape and
the time variability of X-ray \syn\ emission in SNRs \citep[see,
e.g.,][]{Uchiyama_J1713_2007,BUE2008}.

On the other hand, if cascading to short scales is important,
magnetic energy is efficiently transfered to the background gas and the
amplified magnetic field is considerably reduced. The heating of the
shock precursor by the dissipated turbulence will also be
significant. As shown in the bottom panel of Fig.~\ref{fig_ubj}, the
precursor temperature can be orders of magnitude greater with cascading
than without.
The low-energy superthermal particles (up to $p \approx m_p c$) are more
abundant in the presence of cascading and dissipation (top panel of
Figure~\ref{fig_fl}) due to heating-boosted injection
\citep[][]{VBE2008}, which may influence bremsstrahlung and spectral
line emission in SNRs.

\acknowledgments
We thank the anonymous referee for the constructive and useful
comments. We acknowledge support from NASA grants ATP02-0042-0006,
NNH04Zss001N-LTSA, 06-ATP06-21, and RBRF grant 09-02-12080.

\bibliographystyle{apj}

%

\begin{thebibliography}{30}
\expandafter\ifx\csname natexlab\endcsname\relax\def\natexlab#1{#1}\fi

\bibitem[{{Amato} \& {Blasi}(2006)}]{AB2006}
{Amato}, E., \& {Blasi}, P. 2006, \mnras, 371, 1251

\bibitem[{{Bamba} {et~al.}(2003){Bamba}, {Yamazaki}, {Ueno}, \&
  {Koyama}}]{BambaEtal2003}
{Bamba}, A., {Yamazaki}, R., {Ueno}, M., \& {Koyama}, K. 2003, \apj, 589, 827

\bibitem[{{Bell}(2004)}]{Bell2004}
{Bell}, A.~R. 2004, \mnras, 353, 550

\bibitem[{{Bell}(2005)}]{Bell2005}
---. 2005, \mnras, 358, 181

\bibitem[{Bell \& Lucek(2001)}]{BL2001}
Bell, A.~R., \& Lucek, S.~G. 2001, MNRAS, 321, 433

\bibitem[{Berezhko \& Ellison(1999)}]{BE99}
Berezhko, E.~G., \& Ellison, D.~C. 1999, ApJ, 526, 385

\bibitem[{Biskamp(2003)}]{Biskamp2003}
Biskamp, D. 2003, Magnetohydrodynamic turbulence (Cambridge, U.K. ; New York:
  Cambridge University Press), 297

\bibitem[{{Blandford} \& {Eichler}(1987)}]{BE1987}
{Blandford}, R., \& {Eichler}, D. 1987, Phys. Rep., 154, 1

\bibitem[{{Bykov} {et~al.}(2008){Bykov}, {Uvarov}, \& {Ellison}}]{BUE2008}
{Bykov}, A.~M., {Uvarov}, Y.~A., \& {Ellison}, D.~C. 2008, \apjl, 689, L133

\bibitem[{{Diamond} \& {Malkov}(2007)}]{DM2007}
{Diamond}, P.~H., \& {Malkov}, M.~A. 2007, \apj, 654, 252

\bibitem[{{Goldreich} \& {Sridhar}(1995)}]{GS95}
{Goldreich}, P., \& {Sridhar}, S. 1995, \apj, 438, 763

\bibitem[{{Hillas}(2005)}]{Hillas2005}
{Hillas}, A.~M. 2005, J. of Phys. G, 31, 95

\bibitem[{{Jokipii}(1971)}]{Jokipii1971}
{Jokipii}, J.~R. 1971, Reviews of Geophysics and Space Physics, 9, 27

\bibitem[{{Malkov} \& {Diamond}(2009)}]{MD2009}
{Malkov}, M.~A., \& {Diamond}, P.~H. 2009, \apj, 692, 1571

\bibitem[{{Malkov} \& {Drury}(2001)}]{MD2001}
{Malkov}, M.~A., \& {Drury}, L. 2001, Rep. Progr. in Physics, 64, 429

\bibitem[{{Matthaeus} {et~al.}(2009){Matthaeus}, {Oughton}, \&
  {Zhou}}]{Matthaeus2009}
{Matthaeus}, W.~H., {Oughton}, S., \& {Zhou}, Y. 2009, \pre, 79, 035401

\bibitem[{{Palmer}(1982)}]{Palmer1982}
{Palmer}, I.~D. 1982, Reviews of Geophysics and Space Physics, 20, 335

\bibitem[{{Pelletier} {et~al.}(2006){Pelletier}, {Lemoine}, \&
  {Marcowith}}]{PLM2006}
{Pelletier}, G., {Lemoine}, M., \& {Marcowith}, A. 2006, \aap, 453, 181

\bibitem[{{Riquelme} \& {Spitkovsky}(2009)}]{Riquelme2009}
{Riquelme}, M.~A., \& {Spitkovsky}, A. 2009, \apj, 694, 626

\bibitem[{{Toptygin}(1985)}]{Toptygin1985}
{Toptygin}, I.~N. 1985, {Cosmic rays in interplanetary magnetic fields}
  (Dordrecht, D.~Reidel Publishing Co.)

\bibitem[{{Uchiyama} {et~al.}(2007){Uchiyama}, {Aharonian}, {Tanaka},
  {Takahashi}, \& {Maeda}}]{Uchiyama_J1713_2007}
{Uchiyama}, Y., {Aharonian}, F.~A., {Tanaka}, T., {Takahashi}, T., \& {Maeda},
  Y. 2007, \nat, 449, 576

\bibitem[{{Vainshtein} {et~al.}(1993){Vainshtein}, {Bykov}, \&
  {Toptygin}}]{VBT93}
{Vainshtein}, S.~I., {Bykov}, A.~M., \& {Toptygin}, I.~N. 1993, {Turbulence,
  current sheets, and shocks in cosmic plasma} (Gordon \& Breach, New York)

\bibitem[{{Verma}(2004)}]{Verma2004}
{Verma}, M.~K. 2004, Phys. Rep., 401, 229

\bibitem[{{Vink}(2008)}]{Vink2008}
{Vink}, J. 2008, in AIP Conference Proceedings, Vol. 1085, 169--180

\bibitem[{{Vink} \& {Laming}(2003)}]{VL2003}
{Vink}, J., \& {Laming}, J.~M. 2003, \apj, 584, 758

\bibitem[{{Vladimirov} {et~al.}(2006){Vladimirov}, {Ellison}, \&
  {Bykov}}]{VEB2006}
{Vladimirov}, A., {Ellison}, D.~C., \& {Bykov}, A. 2006, \apj, 652, 1246

\bibitem[{{Vladimirov}(2009)}]{Vladimirov_Diss}
{Vladimirov}, A.~E. 2009, Ph.D.~Thesis, NCSU

\bibitem[{{Vladimirov} {et~al.}(2008){Vladimirov}, {Bykov}, \&
  {Ellison}}]{VBE2008}
{Vladimirov}, A.~E., {Bykov}, A.~M., \& {Ellison}, D.~C. 2008, \apj, 688, 1084

\bibitem[{{Zirakashvili} \& {Ptuskin}(2008)}]{ZP2008}
{Zirakashvili}, V.~N., \& {Ptuskin}, V.~S. 2008, \apj, 678, 939

\bibitem[{{Zirakashvili} {et~al.}(2008){Zirakashvili}, {Ptuskin}, \&
  {V{\"o}lk}}]{ZPV2008}
{Zirakashvili}, V.~N., {Ptuskin}, V.~S., \& {V{\"o}lk}, H.~J. 2008, \apj, 678,
  255

\end{thebibliography}
%

\end{document}